\documentclass[12pt]{iopart}

\usepackage{iopams} 

\expandafter\let\csname equation*\endcsname\relax
\expandafter\let\csname endequation*\endcsname\relax
\usepackage{amsmath}
\usepackage{amssymb}
\usepackage{physics}
\usepackage{graphicx}
\usepackage[english]{babel}
\usepackage{bm}
\usepackage{hyperref}
\hypersetup{pdfstartview={FitH},pdfpagemode={UseNone},
            colorlinks,linkcolor=blue, citecolor=blue, urlcolor=blue,
            bookmarksopen=true, pdfnewwindow=true}
\usepackage[all]{hypcap}

\usepackage[labelformat=simple,labelfont=bf,labelsep=period,justification=raggedright]{caption}

\begin{document}

\title[Complete conversion between one and two photons]{Complete conversion between one and two photons in nonlinear waveguides with tailored dispersion}

\author{Alexander S. Solntsev$^1$, Sergey V. Batalov$^{2,3}$, Nathan~K.~Langford$^1$, Andrey A. Sukhorukov$^{4,5}$}
\address{$^1$ School of Mathematical and Physical Sciences, University of Technology Sydney, 15 Broadway, Ultimo NSW 2007, Australia}
\address{$^2$ Institute of Metal Physics, UB RAS, Soﬁa Kovalevskaya str., 18, Ekaterinburg, 620108, Russia}
\address{$^3$ Institute of Physics and Technology, Ural Federal University, Mira str. 19, Ekaterinburg, 620002, Russia}
\address{$^4$ Research School of Physics, Australian National University, Canberra, ACT 2601, Australia}
\address{$^5$ ARC Centre of Excellence for Transformative Meta-Optical Systems (TMOS), Australia}

\def\mailto#1{{\tt \href{mailto:#1}{#1}}}
\eads{\mailto{alexander.solntsev@uts.edu.au}, \mailto{andrey.sukhorukov@anu.edu.au}}

\vspace{10pt}
\begin{indented}
\item[]September 2021
\end{indented}

\begin{abstract}
High-efficiency photon-pair production is a long-sought-after goal for many optical quantum technologies, and coherent photon conversion processes are promising candidates for achieving this.
We show theoretically how to control coherent conversion between a narrow-band pump photon and broadband photon pairs in nonlinear optical waveguides by tailoring frequency dispersion for broadband quantum frequency mixing.
We reveal that complete deterministic conversion as well as pump-photon revival can be achieved at a finite propagation distance.
We also find that high conversion efficiencies can be realised robustly over long propagation distances.
These results demonstrate that dispersion engineering is a promising way to tune and optimise the coherent photon conversion process.
\end{abstract}

%
%
%
%
%

\section{Introduction}

Optical nonlinearities play a vital role in the development of quantum-enhanced technologies based on quantum optics and photonic quantum information~\cite{Knill:2001-46:NAT, Kok:2007-135:RMP, Schnabel:2010-121:NCOM, Bartlett:2002-97904:PRL}.
Typically realised in the optical regime through interactions with individual atomic systems~\cite{Turchette:1995-4710:PRL, Volz:2014-965:NPHOT, Semiao:2005-64305:PRA} or atomic media~\cite{Beck:2016-9740:PNAS, Tiarks:2016-e1600036:SCA, Ou:1988-50:PRL, Fiorentino:2002-983:IPTL, Reim:2010-218:NPHOT, Fabre:1994-1337:PRA}, such nonlinearities are very weak at the few-photon level unless enhanced by, e.g., cavity confinement~\cite{Hood:1998-4157:PRL, Groblacher:2009-724:NAT}, coherent ensemble effects~\cite{Julsgaard:2004-482:NAT, Beck:2016-9740:PNAS, Tiarks:2016-e1600036:SCA} or strong classical pump fields~\cite{Vandevender:2004-1433:JMO, Reim:2010-218:NPHOT}.
In 2001, the pioneering KLM proposal~\cite{Knill:2001-46:NAT} demonstrated that direct photon-photon nonlinearities could be circumvented using the strong local nonlinearity provided by avalanche photodetection, with feedforward and teleportation
, to enable efficient, fault-tolerant quantum computation with linear optics~\cite{Kok:2007-135:RMP}.
While subsequent proposals have greatly reduced the significant physical resource overheads of a linear optics approach~\cite{Nielsen:2004-40503:PRL, Browne:2005-10501:PRL, Nickerson:1810.09621:ARXIV}, nonlinear optical quantum computing~\cite{Munro:2005-33819:PRA, Nemoto:2004-250502:PRL, Franson:2004-62302:PRA, Langford:2011-360:NAT} still provides an enticing alternative, with strong, direct photon-photon nonlinearities promising to minimise the intensive resource requirements entailed by probabilistic interactions.

After initial proposals~\cite{Munro:2005-33819:PRA, Nemoto:2004-250502:PRL, Franson:2004-62302:PRA} based on photon-photon cross-phase modulation, developments in nonlinear optical quantum computing slowed for some time after several theoretical ``no-go'' theorems (e.g.,~\cite{Shapiro:2006-62305:PRA, Gea-Banacloche:2010-43823:PRA}).
These suggested the goal of directly realising multiphoton gates between travelling photon pulses using optical nonlinearities faced fundamental roadblocks, due to finite-bandwidth and spectral entanglement effects that create a trade-off between nonlinear interaction strengths and gate fidelities.
Later, Langford \emph{et al.} introduced a new photon-level nonlinear process, coherent photon conversion (CPC), that could sidestep these issues and provide a versatile building block for a new, scaleable quantum computing architecture~\cite{Langford:2011-360:NAT}.
Generalising well-established concepts in spontaneous parametric downconversion and single-photon up-conversion~\cite{Ou:1988-50:PRL, Vandevender:2004-1433:JMO, Koshino:2009-13804:PRA}, CPC provides a nonlinear module that enables deterministic multiphoton gates, high-quality heralded single- and multiphoton states free from higher-order imperfections, and robust, high-efficiency detection.
Initial analysis suggested similar fundamental roadblocks also limit the performance of CPC-based operations for pulsed (broadband) photons~\cite{Viswanathan:2015-42330:PRA}.
Subsequent work, however, including by the authors of Ref.~\cite{Viswanathan:2015-42330:PRA}, has identified operational paradigms for CPC and other nonlinear optical quantum operations that potentially circumvent these issues~\cite{Xia:2016-23601:PRL, Brod:2016-80502:PRL, Viswanathan:2018-32314:PRA, Chudzicki:2013-42325:PRA, Niu:2018-160502:PRL}, for example by using dispersion and group-velocity engineering to control the pulse interactions~\cite{Xia:2016-23601:PRL, Mosley:2008-133601:PRL, Halder:2009-4670:OE} or atom-mediated pulse interactions~\cite{Brod:2016-80502:PRL, Koshino:2009-13804:PRA}.
Despite these promising steps, however, it remains an important open question to identify the ultimate limitations to high-efficiency CPC operations.

Spontaneous parametric down-conversion (SPDC), the major workhorse for nonclassical light sources in quantum optics and quantum information experiments, has found important application in two different operating paradigms.
When implemented with pulsed pump lasers, SPDC produces more narrowband, possibly factorable photon pairs, enabling the synchronised, multipair photon production that is vital for quantum computing~\cite{Lanyon:2007-250505:PRL, Walther:2005-169:NAT, Solntsev:2017-19:RPH, Zhong:2018-250505:PRL}, quantum networking~\cite{Duan:2001-413:NAT, Barrett:2005-60310:PRA, Nunn:2013-133601:PRL} and quantum sensing~\cite{Thomas-Peter:2011-55024:NJP, Xiang:2011-43:NPHOT} applications.
When implemented with narrowband pump lasers, SPDC leads to highly broadband, spectrally entangled output photons, which can be important for quantum communication, imaging and sensing tasks~\cite{Humphreys:2014-130502:PRL, Brecht:2015-41017:PRX, Barreiro:2005-260501:PRL, Erkmen:2010-405:ADOP, Solntsev:2018-21301:APLP}.

To date, experimental and theoretical investigations of CPC have mostly focused on the broadband pump paradigm.
In this regime, experimental demonstrations include conversion of a single photon to two in an optical fiber~\cite{Langford:2011-360:NAT}, and two to one in a nonlinear waveguide~\cite{Guerreiro:2014-173601:PRL}. While the demonstrated conversion efficiency was relatively low in these first experiments, it can be improved by orders of magnitude by using strongly nonlinear materials and optimizing the waveguide geometry~\cite{Dot:2014-43808:PRA}, based on solutions derived in the low-conversion regime.
In the narrowband-pump regime, it has also been predicted that complete conversion of a single pump photon to two down-converted photons can be achieved in a nonlinear waveguide with quadratic frequency dispersion~\cite{Antonosyan:2014-22:OC, Yanagimoto:2009.01457:ARXIV}.

In this work, we show how dispersion engineering can be used to tune the photon-conversion process.
This addresses a key open goal in this context, which is to study the optimal conditions for achieving deterministic photon conversion both from a single photon to a pair and also backward.
We study a range of dispersion scenarios and show that we can reach 100\% forward and backward conversion efficiency at a finite propagation length.
We also show that it is possible to realise robust conversion between one and two photons, where high conversion efficiencies can be realised over a large propagation distance range.
These are nontrivial results due to the complex dynamics involving the one- and two-photon states across a broad optical frequency spectrum.









\section{Model}
\subsection{Conceptual framework}

\begin{figure}[t]
	\centering
	\includegraphics[width=11.5cm]{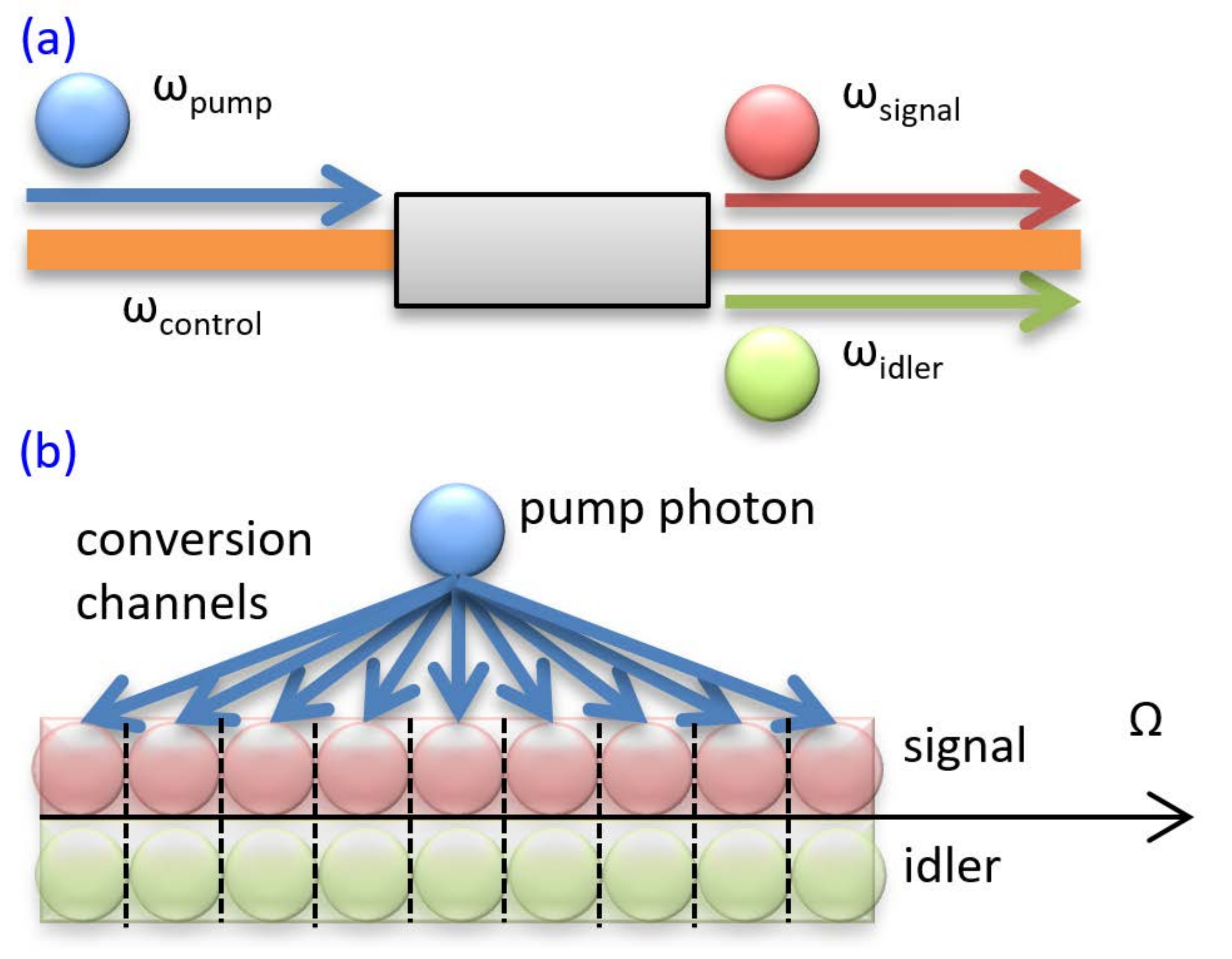}%
	\caption{(a) The concept of coherent conversion of a pump photon into signal and idler photons with frequencies $\omega_{s}$ and $\omega_{i}$, respectively.
(b) Photon conversion channels associated with different frequency detunings between the signal and idler photons. Utilising a larger number of coherent photon conversion channels by increasing the generated photon-pair bandwidth can drastically increase the efficiency of the process.}
\label{Fig:intro}
\end{figure}

We start by considering the coherent conversion of a pump photon with a central frequency $\omega_{p}$ into signal and idler photons, see Fig.~\ref{Fig:intro}(a). 
This process can be realized 
in media with cubic nonlinearity through four-wave mixing involving a high-power control wave at a different frequency $\omega_{c}$~\cite{Langford:2011-360:NAT}. For narrow-band spectra of the input photon and the control wave, the signal (index $s$) and idler (index $i$) frequencies are related due to the energy conservation as $\omega_{s} + \omega_{i} = \omega_{p} + \omega_{c}$. On the other hand, the splitting between the two photon frequencies can be arbitrary, $\omega_{s} - \omega_{i} = \Omega$. 
%
%
In the following, we show how to utilize a larger number of coherent photon conversion channels corresponding to different detunings $\Omega$ [Fig.~\ref{Fig:intro}(b)] in order to reach complete photon conversion in the forward and backward directions.
 
The dynamics of the coherent photon conversion is determined by the phase mismatch of the four-wave-mixing process across a range of photon frequencies, $\Delta\beta(\Omega) = \beta(\omega_{s}) + \beta(\omega_{i}) - \beta(\omega_{p}) - \beta(\omega_{c})$, where $\beta$ is the propagation constant of the waveguide mode at the corresponding optical frequency. 
Whereas it is generally accepted that most efficient conversion occurs in the regime of phase matching~\cite{Dot:2014-43808:PRA}, the shape of the mismatch dependence around the phase-matching point plays a critically important role in the high-conversion regime. We consider coupled equations for the single and two-photon wavefunctions in the strong conversion regime~\cite{Antonosyan:2014-22:OC}, and establish their mathematical equivalence to the fundamentally important phenomenon of the decay of a discrete atomic state to a continuum~\cite{Akulin:2014:DynamicsComplex}, however the role of temporal evolution is replaced by the propagation distance along the waveguide ($z$). 
Therefore, by choosing a particular waveguide length, we can access any intermediate stage of the decay dynamics.

\subsection{Theoretical model}

Now let us derive a model following Ref.~\cite{Dot:2014-43808:PRA} and restricting four-wave mixing to only one polarization component in a single-mode waveguide. In this case the propagating field can be written in the form
\begin{align}
    \hat{E}(z,t) = \sqrt{\frac{\hbar}{4\pi \varepsilon_0 c A_{\text{eff}}}} \int_0^{\infty} d\omega \sqrt{\frac{\omega}{n(\omega)}}
    \hat{a}(\omega, z) e^{-i\omega t} + \text{H.c.},
    \label{eq:E}
\end{align}
where effective area of the waveguide mode $A_{\text{eff}}$ is taken to be the same for all
frequency components in the waveguide. Length $L$ of the fiber is assumed to be large enough
for the continuous limit to be valid. Note that operators $\hat{a}(\omega, z)$
in Eq.~\eqref{eq:E} are chosen to be dimensional with units of $(\delta\omega)^{-1/2}$
\cite{Blow:1990-4102:PRA}, where $\delta\omega$ is the frequency spacing due to
periodic boundary conditions $\delta\omega={2\pi}/{T}$ and $T=L/c$ is the quantization time.

In absence of nonlinearity, evolution of operators $\hat{a}$ can be found \cite{Huttner:1990-5594:PRA} 
to be $\hat{a}(\omega, z) = \hat{a}_0 \, e^{i \beta(\omega) z}$,
where $\beta(\omega) = n(\omega) \omega / c$. In the nonlinear medium operators $\hat{a}_0$
become functions of coordinate and frequency.
If the strong control wave can be taken as classical ($\hat{a}_0(\omega_{c}, z) \equiv A_{c}(z)$)
and undepleted ($|A_{c}(z)|^2 = |A_{c}(0)|^2$), then the evolution of the weak single-photon pump
and generated signal and idler modes is governed by the following set of equations
\cite{Dot:2014-43808:PRA} \footnote{Equation (17) in \cite{Dot:2014-43808:PRA} contains a typo.
Factor $2\pi/T$ should be replaced by $T/2\pi$.}
\begin{align}
    \frac{\partial \hat{a}_0(\omega_{p}, z)}{\partial z} &= 2 i \gamma 
        \sqrt{P_c\zeta_p} \, \frac{T}{2\pi} \int d\omega_s \hat{a}_0(\omega_s, z) \,
            \hat{a}_0(\omega_{c} + \omega_{p} - \omega_{s}, z) e^{i(\Delta k - \gamma\,P_c) z} \notag \\ 
       & + 2 i \gamma P_c \hat{a}_0(\omega_{p}, z) ,
    \label{eq:weak_pump} \\
    \frac{\partial \hat{a}_0(\omega_s, z)}{\partial z} &= 2 i \gamma 
        \sqrt{P_c\zeta_p} \, \hat{a}^\dagger_0(\omega_{c} + \omega_{p} - \omega_s, z) \,
            \hat{a}_0(\omega_{p}, z) e^{-i(\Delta k - \gamma\,P_c) z} \notag \\
       & + 2 i \gamma P_c \hat{a}_0(\omega_s, z) .
    \label{eq:generated}
\end{align}
Here $\gamma(\omega_s) = {3 \chi^{(3)} \omega_s}/[2\varepsilon_0 c^2 n^2(\omega_s) A_{\text{eff}}]$
is the standard waveguide parameter.
We will assume that 
the value of $\gamma$ is approximately the same for all the frequency modes \cite{Agrawal:2013:NonlinearFiber}.
Parameter $P_c = {2\pi \hbar \omega_{c}}{T^{-2}}\;|A_{c}|^2$ measures the peak power of
the strong control wave. Parameter $\zeta_p = {2\pi \hbar\omega_{p}}{T^{-2}}$ is defined so
that $P_p = \zeta_p\times \langle a^{\dagger}(\omega_{p}, 0) a(\omega_{p}, 0)\rangle$ is
the photon population of the weak pump at the entrance of the medium. In case of a single-photon
pump we have $P_p = \zeta_p\times {T}/(2\pi)$. $\Delta k$ is a phase mismatch based on the waveguide geometry. The total phase mismatch  $\Delta\beta = - \left(\Delta k + \gamma \, P_c\right)$ can be modified dynamically by changing the strong pump power.

Now let us introduce the function for the pump photon dynamics
\begin{align}
    U(z) = e^{2 i \gamma P_c z} \,
    \frac{2\pi}{T} \bra{0} \hat{a}_0(\omega_{p}, 0) \hat{a}^{\dagger}_0(\omega_{p}, z) \ket{0}.
    \label{eq:U}
\end{align}
This is the probability amplitude to find a weak pump photon at distance $z$ provided that
there was one photon at distance $z=0$. Factor $2\pi/T$ accounts for units of operators $\hat{a}_0$
and exponential factor is introduced for convenience. Taking coordinate derivative
we obtain the following equation
\begin{align}
    \frac{dU}{dz} = -i \, \chi \, \frac{T}{2\pi} \int d\omega_s \, V(\omega_s, z),
    \label{eq:dU}
\end{align}
where $\chi = 2 \gamma \, \sqrt{P_c P_p}$.
New quantity $V(\omega, z)$ in Eq.~\eqref{eq:dU} is defined as follows
\begin{align}
\begin{split}
    V(\omega_s, z) & = \left(\frac{2\pi}{T}\right)^{3/2}
    e^{i(3\gamma P_c -\Delta k) z}  \\
    & \times \bra{0}
        \hat{a}_0(\omega_{p}, 0) \,
        \hat{a}_0^{\dagger}(\omega_{c}+\omega_{p} - \omega_s, z) \,
        \hat{a}_0^{\dagger}(\omega_s, z)
    \ket{0},
\end{split}
\end{align}
Its physical meaning is the probability amplitude to find
a pair of photons $\omega_s$ and $\omega_i=\omega_c+\omega_p-\omega_s$ at
distance $z$, provided that there was a single pump photon with frequency
$\omega_{p}$ at distance $z=0$. Differentiating $V(\omega_s, z)$ with respect
to $z$ we get a second equation
\begin{align}
    \frac{dV}{dz} = -i \Delta \beta  V - i \, \chi \, U,
    \label{eq:dV}
\end{align}
where we took into account that $\hat{a}_0(\omega, z)\ket{0} \equiv 0$ and
$\hat{a}_0(\omega, z) \,\hat{a}^{\dagger}_0(\omega, z) \ket{0} \equiv {T}/(2\pi) \ket{0}$.
Equations \eqref{eq:dU} and \eqref{eq:dV} together with ``initial conditions''
$U(0) = 1$, $V(\omega, 0) = 0$ can be used to describe dynamics of our system, effectively representing solutions of the operator equations \eqref{eq:weak_pump} and \eqref{eq:generated}. The temporal dynamics for the photon pair packet can then be calculated via Fourier transform $\tilde{V}(\tau, z) =(2 \pi)^{-1/2} \int_{-\infty}^{\infty} \! U(\omega, z) \mathrm{e}^{i\omega \tau}\, d\omega $. 
We note that in the framework of Eqs.~(\ref{eq:dU}) and~(\ref{eq:dV}), the combined population of pump photons $I_p$ and signal-idler pairs $I_s$ is conserved,
\begin{equation}
    \frac{d}{dz}\left[I_p(z) + I_s(z)\right] = 0,
\end{equation}
where
\begin{equation}
    I_p(z) = |U(z)|^2, \quad 
          I_s(z) = \int d \omega_s |V(\omega_s,z)|^2 
                 = \int d \tau |\tilde{V}(\tau, z)|^2.
\end{equation}
With no loss of generality, in the numerical examples below we consider a normalization of variables such that $\chi T / (2\pi) = 1$. We also introduce a notation $\Omega = \omega_s - \omega_{s0}$, where $\omega_{s0}$ is a characteristic signal photon frequency.

\begin{figure}[t]
	\centering
	\includegraphics[width=11.5cm]{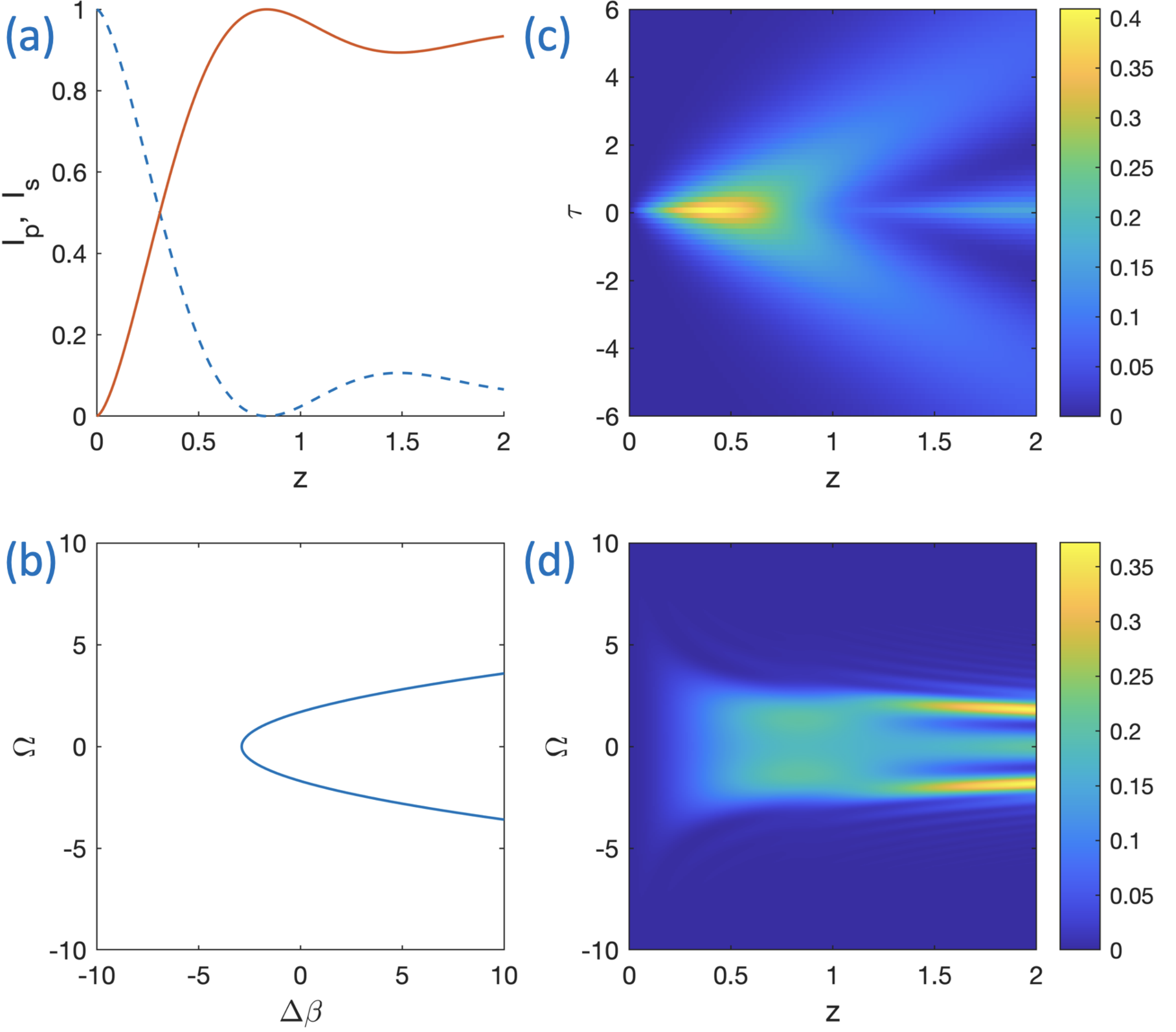}%
	\caption{
	Complete conversion of one photon into two at a finite propagation distance ($z_c=0.83$).
	(a) Pump (dashed blue) and signal / idler (solid red) photon populations vs. the propagation distance $z$ in the case of quadratic waveguide dispersion $\Delta \beta = \Omega^2 - 2.88$ shown in (b). (c) Temporal and (d) spectral dynamics of the biphoton population $|V|^2$ vs.the propagation distance $z$.
	}
	\label{Fig:quadratic}
\end{figure}

\section{Results}

\subsection{Complete conversion of one photon into two}

Now we analyze the effects of frequency dispersion $\Delta\beta(\Omega)$ on the photon dynamics.
A linear dispersion corresponds to Markovian decay and allows 100\% conversion efficiency only in the limit of infinite propagation distance $z \rightarrow \infty$. In contrast, the higher-order dispersion enables complete conversion. 
Specifically, 100\% conversion efficiency [Fig.~\ref{Fig:quadratic}(a)] at a finite propagation distance in normalized units $z_c=0.83$ can be enabled by the shifted quadratic dispersion $\Delta \beta = \Omega^2 - 2.88$  [Fig.~\ref{Fig:quadratic}(b)]. We find that there appear multiple parameter regions corresponding to the full conversion, see \ref{appendix}, whereas only specific cases we identified in previous studies~\cite{Antonosyan:2014-22:OC, Yanagimoto:2009.01457:ARXIV}. We note that the mathematical model of complete conversion under quadratic dispersion is equivalent to a decay of atomic state to a continuum near a photonic band-edge~\cite{John:1994-1764:PRA}, and accordingly the effect of complete transitional decay can also happen for atoms, see \ref{appendix} for details.   
Interestingly, after the rebound from complete conversion, in the temporal domain an entangled state with an approximate form $\ket{0}_s \ket{0}_i$ + $\ket{\tau_0}_s \ket{-\tau_0}_i$ + $\ket{-\tau_0}_s \ket{\tau_0}_i$ is generated, where $\tau$ defines delay or advance for the signal/idler photons in a moving frame. 
Specifically, the signal and idler photons are either travelling together in the central peak or with a time delay $\pm \tau_0$ linearly growing with propagation distance $z$ [Fig.~\ref{Fig:quadratic}(c)]. A similar state with three peaks is also formed in the spectral domain, although the distance between the peaks remains constant with increasing $z$ [Fig.~\ref{Fig:quadratic}(d)].
The considered quadratic dispersion typically occurs near the degeneracy point when $\omega_s \approx \omega_i$~\cite{Solntsev:2012-27441:OE}, which can be readily accessed experimentally.

\begin{figure}[t]
	\centering
	\includegraphics[width=11.5cm]{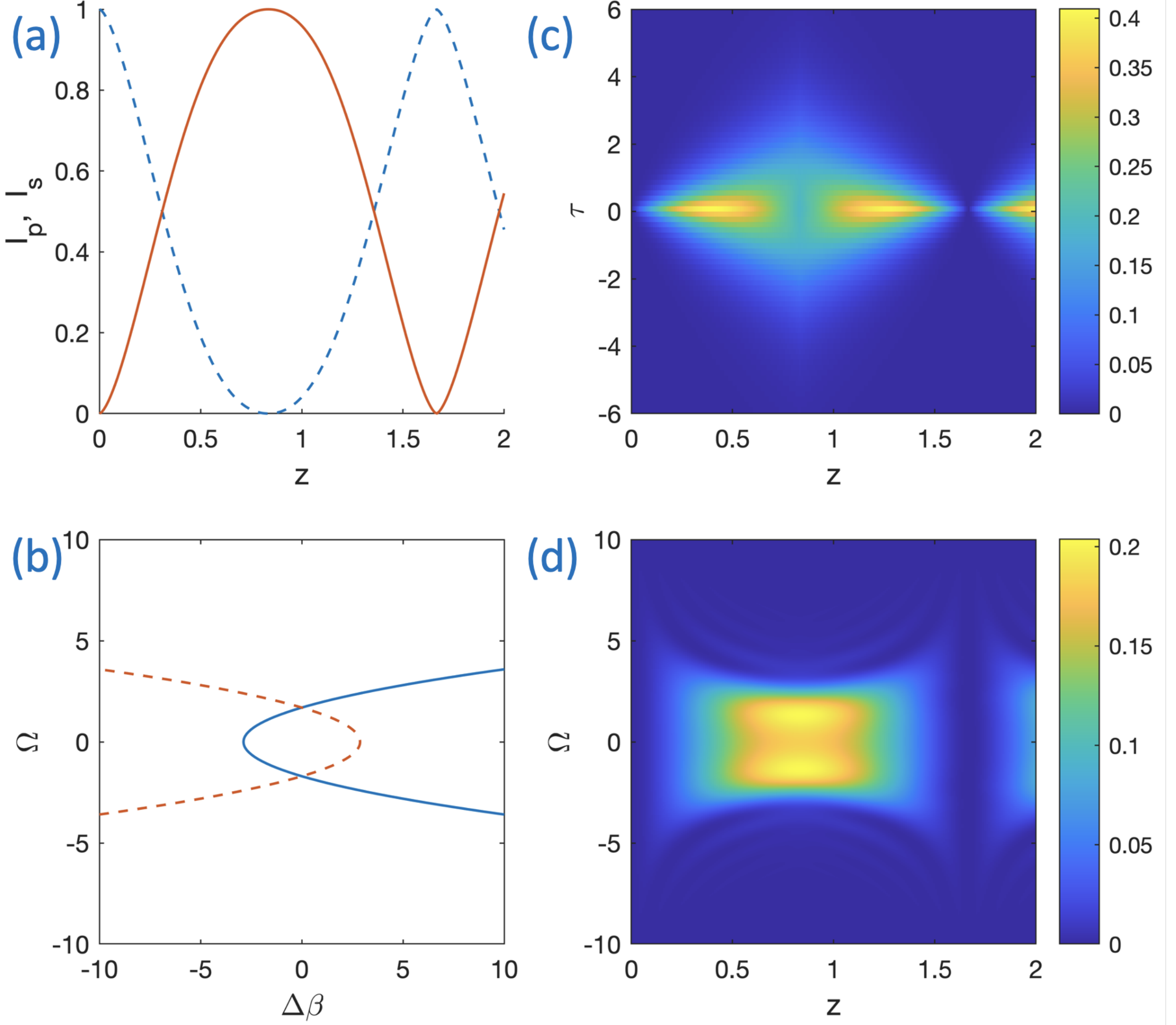}%
	\caption{
	Forward and backward conversion between one- and two-photon states achieved by reversing the sign of the dispersion. 
	(a) Pump (dashed blue) and signal / idler (solid red) photon populations vs. the propagation distance $z$. (b) Quadratic waveguide dispersion $\Delta \beta = \Omega^2 - 2.88$ for $z \le z_{\rm inv} \simeq 2.53$ (solid blue line) and the specially inverted dispersion $\Delta \beta = - \Omega^2 + 2.88$ for $z > z_{\rm inv}$ (dashed red). 
	(c)~Temporal and (d)~spectral dynamics of the biphoton population $|V|^2$ vs.the propagation distance $z$. 
	}
	\label{Fig:reversal}
\end{figure}

\subsection{Photon conversion reversal}

A regime which is of particular interest in the context of photonic quantum computing is where coherent conversion can be realised in both forwards and backwards directions, allowing a pair of photons to also completely convert back into one photon, similar to Rabi oscillations~\cite{Langford:2011-360:NAT, Dot:2014-43808:PRA}.
However, when dispersion is present, this regime is no longer possible in homogeneous waveguides.
This is because the spectral distribution of the generated pair of photons interacts in a complex way with the nontrivial waveguide dispersion, and does not produce full coherent recombination in the backwards process~\cite{Antonosyan:2014-22:OC, Yanagimoto:2009.01457:ARXIV}.


\begin{figure}[t]
	\centering
	\includegraphics[width=13.5cm]{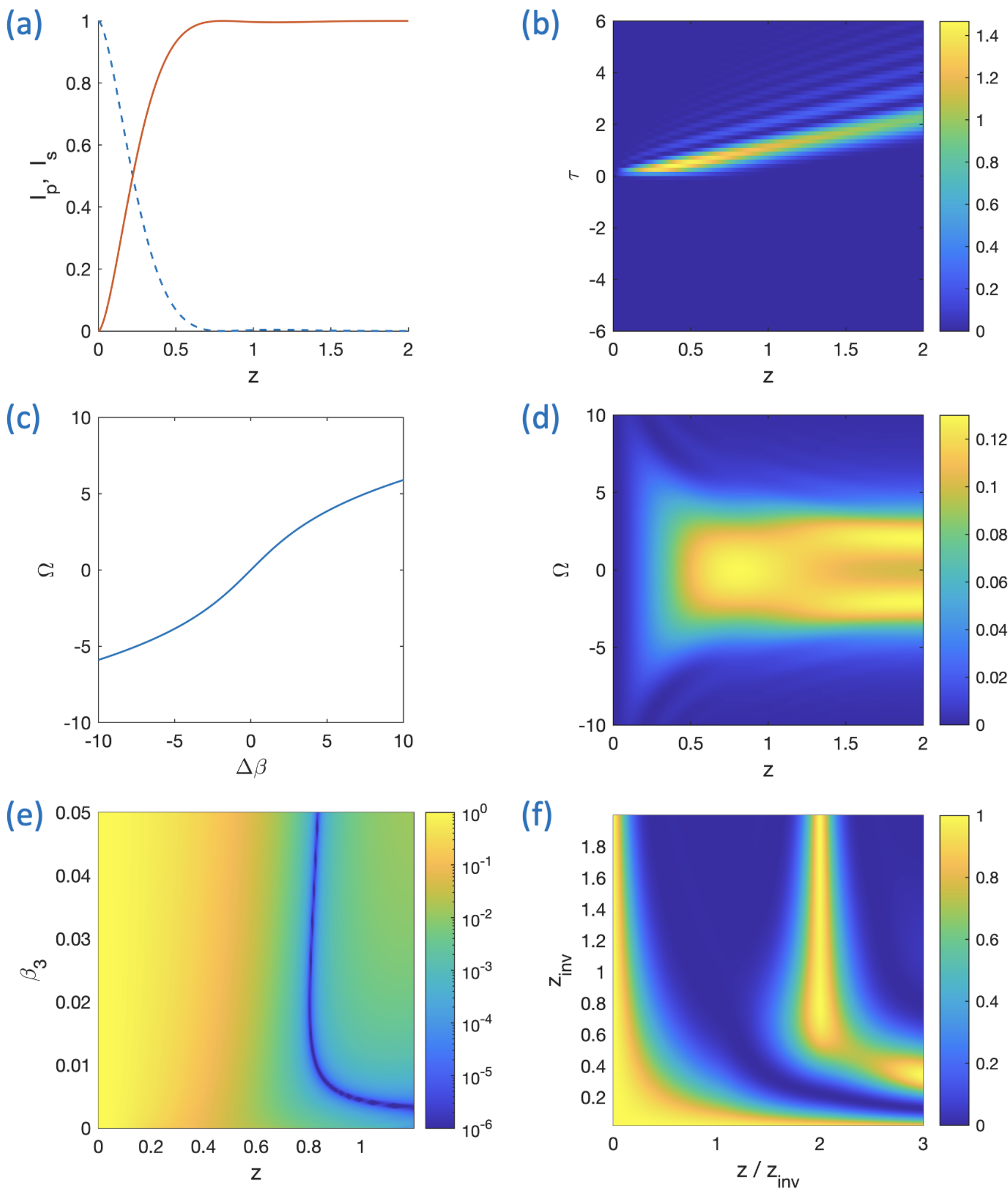}%
	\caption{
	Robust conversion between one and two photons achieved through engineering higher-order frequency dispersion.
	(a) Pump (dashed blue) and signal / idler (solid red) photon populations vs. the propagation distance $z$ in the case of cubic waveguide dispersion. (b) Temporal dynamics of the biphoton population $|V|^2$ vs.the propagation distance $z$. (c) Dispersion $\Delta \beta = \beta_3 \, \Omega^3 + \Omega$, with $\beta_3 = 0.02$. (d) Spectral dynamics of the biphoton population $|V|^2$ vs. the propagation distance $z$. (e) Pump photon population $|U|^2$ shown in logarithmic scale vs. the propagation distance $z$ and third-order dispersion coefficient $\beta_3$. (f) Pump photon population $|U|^2$ vs. the coordinate of dispersion sign inversion $z_{\rm inv}$ and the normalized distance $z/z_{\rm inv}$ for $\beta_3 = 0.02$.
	}	
	\label{Fig:cubic}
\end{figure}

Here, however, we show that complete forward and backward conversion between one- and two-photon states can be achieved through waveguide dispersion engineering. 
Specifically, by inverting the sign of the dispersion $\Delta \beta \rightarrow -\Delta \beta$ at the point of maximum conversion, the photon population dynamics reverses according to the symmetry of the governing equations. Thereby, at twice the complete forward conversion distance, a biphoton converts back into a single pump photon [Fig.~\ref{Fig:reversal}(a)]. In this case, mirroring the dispersion at $z_{\rm inv} \simeq 2.53$ [Fig.~\ref{Fig:reversal}(b)] is sufficient, which can be achieved through a tailored waveguide engineering. The biphoton wavepacket in the temporal domain [Fig.~\ref{Fig:reversal}(c)] as well as its spectrum [Fig.~\ref{Fig:reversal}(d)] also show complete reversal. During this process, the phase of the original photon is shifted by $\pi$, in a similar way to the nonlinear optical control-phase gates realised in~\cite{Langford:2011-360:NAT}. 
Given the complex spectral dynamics that takes place during coherent photon conversion in the presence of nontrivial waveguide dispersion, it is already interesting to observe that complete reversal is still possible, and surprising that it can be achieved with such a mathematically simple dispersion modification.

\subsection{Robust photon conversion mediated by higher-order dispersion}

We then find that more robust operation of photon conversion can be achieved by further tailoring the higher-order waveguide dispersion.
As discussed above, in the case of quadratic dispersion, the pump photon population quickly rebounds after complete conversion, which would require highly precise optimisation of the frequency dispersion and optical nonlinearity in experiments.
We find that conversion with strongly reduced sensitivity to experimental inaccuracies can be achieved in waveguides with engineered cubic frequency dispersion. We show an example of 100\% conversion efficiency at $z_c \simeq 0.8$ in Fig.~\ref{Fig:cubic}(a), followed by an extended region of over 99.5\% conversion. This flat behaviour with respect to an increase of the propagation distance $z$ indicates high robustness against the experimental deviations.
In terms of temporal dynamics, the biphoton shows very limited spreading with a moderate, extended tail [Fig.~\ref{Fig:cubic}(b)], when a tailored cubic dispersion with $\Delta \beta = \beta_3 \, \Omega^3 + \Omega$ with $\beta_3 = 0.02$ is utilized [Fig.~\ref{Fig:cubic}(c)]. In the spectral domain, there are two merged peaks [Fig.~\ref{Fig:cubic}(d)]. 
We show the effect of the third-order dispersion strength on the pump photon evolution in Fig.~\ref{Fig:cubic}(e). We observe that for $\beta_3 \simeq 0.02$, the zero position $z_c$ effectively does not depend on variations of $\beta_3$, indicating 
robustness with respect to $\beta_3$ variations.
Importantly, the robustness of back-conversion from two photons to one photon based on a general approach formulated above, where the dispersion sign is inverted at $z_{\rm inv}$, can be also enhanced by third-order dispersion optimization. We show in Fig.~\ref{Fig:cubic}(f) that for a large range of $z_{\rm inv} \ge z_c \simeq 0.8$, there is nearly complete back-conversion with $|U|^2 \rightarrow 1$ at $z = 2\; z_{\rm inv}$.



\section{Conclusions}

In conclusion, we have shown that complete deterministic conversion between one and two photons can be achieved in nonlinear waveguides with specially engineered frequency dispersion. 
In particular, quadratic dispersion can facilitate 100\% conversion efficiency between one and two photons in the forward and backward directions at finite propagation lengths, 
allowing the pump photon to complete a full oscillation. 
Furthermore, specially optimized cubic dispersion enables highly robust photon conversion, with strongly reduced sensitivity to potential experimental inaccuracies.
This work shows that dispersion can be designed for the high-efficiency production of spectrally entangled, broadband photon pairs, with no higher-order multiphoton terms, which may provide significant benefits for use in advanced quantum communication technologies.



\section*{Acknowledgements}

The authors acknowledge the support by the Australian Research Council (DE180100070, DP160100619, DP190100277).
NKL is funded by the Australian Research Council Future Fellowship  (FT170100399).
Batalov S.V. acknowledges support by the Ministry of Education and Science of the Russian Federation (the theme ``Quantum'', No. AAAA-A18-118020190095-4).
\appendix
\section{Asymptotic analysis of photon populations} \label{appendix}

Let us explore the quadratic dispersion approximation to the total phase mismatch $
\Delta\beta =  \Delta\beta_0 + \beta_2\,\Omega^2
$ in more detail.
Applying Laplace transform $f(s) = \int_0^{\infty} f(z) e^{-s z} dz$ to equations \eqref{eq:dU}, \eqref{eq:dV}
and taking into account conditions at the left waveguide boundary $U(0) = 1$, $V(\Omega, 0) = 0$ we find
\begin{align}
	U(s) = \left[{s + \frac {\alpha \sqrt{i}} {\sqrt{s - i \Delta\beta_0}}}\right]^{-1},
	\label{eq:laplace}
\end{align}
where $\alpha = {\pi \chi^2}/{\sqrt{\beta_2}}$.
The inverse Laplace transform then yields
\begin{align}
	U(z) = e^{i \Delta\beta_0 z} \sum_k \frac {e^{p_k^2 z}} {3 + i \Delta\beta_0 / p_k^2} \left(1 + \erf({p_k \sqrt{z})} \right),
	\label{eq:solution}
\end{align}
where $\erf(x)$ is the error function and $p_k$ are the three roots of the following cubic equation:
\begin{align*}
	p^3 + i \Delta\beta_0 \, p + \alpha \sqrt{i} = 0.
\end{align*}
Roots $p_k$ can be expressed using Vieta's formula as
\begin{align}
	p_k &= \alpha^{1/3} \sigma_k(\gamma), \quad \sigma_k(\gamma) =
    	e^{-\frac{i \pi}4} \left(\zeta^k A  + \zeta^{2k} B \right), \label{eq:roots}\\
	A &=  \sqrt[3]{\frac 1 2 + \frac 1 2 \sqrt{1 - \frac {4\gamma^3}{27}}},
    \quad B = \frac{\gamma} {3 A}, \quad k=0,1,2,  \nonumber
\end{align}
where $\gamma = {\Delta\beta_0} / {\alpha^{2/3}}$ and $\zeta = e^{{2\pi i} / 3}$
is one of the cubic roots of unity. It is convenient to introduce new spatial variable $\xi=\alpha^{2/3} z$,
then solution in Eq.~\eqref{eq:solution} depends on single parameter~$\gamma$:
\begin{align}
	U(\xi, \gamma) = e^{i \gamma \xi} \sum_k \frac {e^{\sigma_k^2 \xi}} {3 + i \gamma / \sigma_k^2} \left(1 + \erf({\sigma_k \sqrt{\xi})} \right),
	\label{eq:solution2}
\end{align}
where $\sigma_k$ are given by Eq.~\eqref{eq:roots} and are functions of $\gamma$ only.

\begin{figure}[t]
	\centering
	\includegraphics[width=13.5cm]{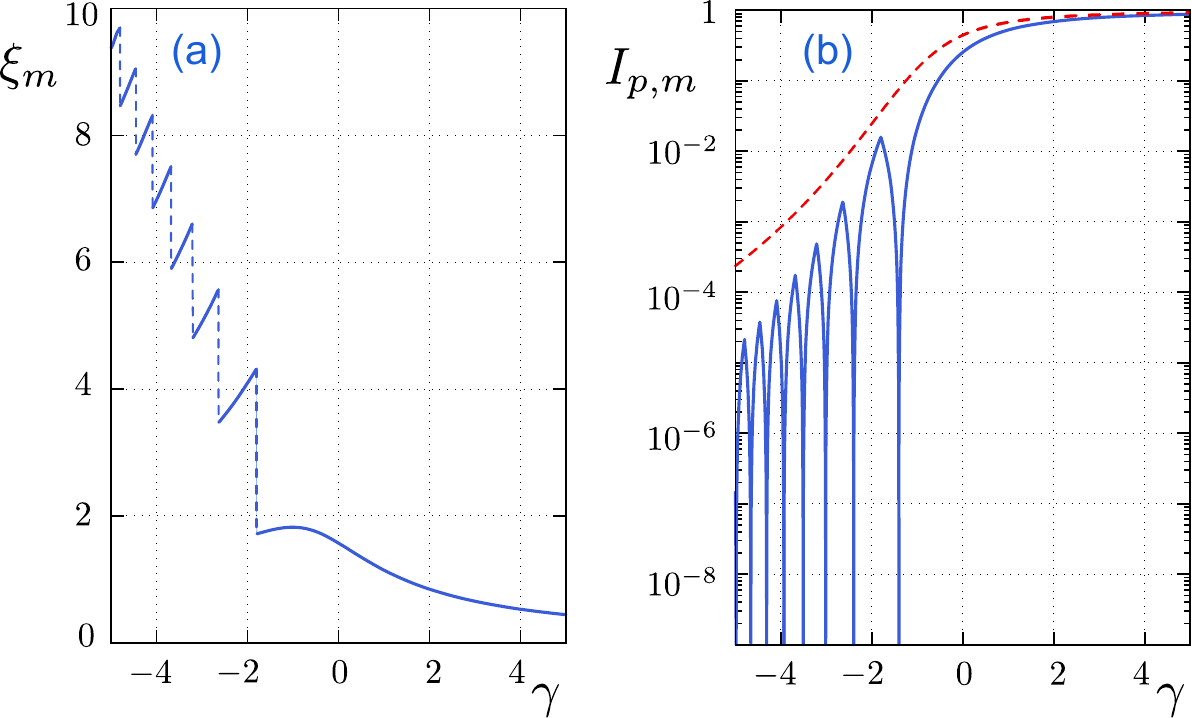}%
	\caption{\label{fig:1}
	(a)~Scaled optimal waveguide length $\xi_m(\gamma)$ corresponding
	to the global minimum of pump photon population.
	Dashed lines mark discontinuities due to changes of local minima.
	(b)~Pump photon population at optimal waveguide lengths $I_{p,m}(\gamma)=|U(\xi_m(\gamma), \gamma)|^2$.
	Red dashed line in (b) corresponds to pump photon population at infinite distance according to Eq.~\eqref{eq:pump_at_infinity}.}
	\end{figure}

Let us calculate the asymptotic value of the pump wave population at infinite distance.
The asymptotics of error function at $z\to\infty$ is given by \cite{OldeDaalhuis:1995-1469:SAM}:
\begin{align*}
	\erf{z} \sim \begin{cases}
		1 - {e^{-z^2}}/({z\sqrt{\pi}}), & z\in\Omega_1,\\
		-1 - {e^{-z^2}}/({z\sqrt{\pi}}), & z\in\Omega_2,\\
		- {e^{-z^2}}/({z\sqrt{\pi}}), & z\in\Omega_3,
	\end{cases}
\end{align*}
where $\Omega_1$, $\Omega_2$ and $\Omega_3$ are sectors of complex plane with
$-\pi / 4 < \arg{z} < \pi / 4$,
${3\pi} / 4 < \arg{z} < {5 \pi} / 4$
and ${\pi} / 4 < \arg{z} < {3 \pi} / 4 \cup {5\pi} / 4 < \arg{z} < {7 \pi} / 4$ respectively.
As $\gamma$ changes from $-\infty$ to $\gamma^{*}=3/{4^{1/3}}$, roots $\sigma_k(\gamma)$ move in complex plane so that
$\sigma_1(\gamma)\in \Omega_3$ and $\sigma_2(\gamma)\in\Omega_2$.
For $\xi\to+\infty$ the contributions from these two roots vanish:
\begin{align}
	e^{\sigma_1^2 \xi}\left(1 + \erf{\sigma_1 \sqrt{\xi}}\right) &\sim e^{\sigma_1^2 \xi} - \frac 1 {\sigma_1 \sqrt{\pi\xi}} \to 0, \label{eq:sigma_1_asymp} \\
	e^{\sigma_2^2 \xi}\left(1 + \erf{\sigma_2 \sqrt{\xi}}\right) &\sim -\frac 1 {\sigma_2 \sqrt{\pi\xi}} \to 0. \label{eq:sigma_2_asymp}
\end{align}
In Eq.~\eqref{eq:sigma_1_asymp} we used the fact that $\Re\left(z^2\right) < 0$ in $\Omega_3$ and
the corresponding exponent rapidly decreases.
The only term surviving at large distances comes from $\sigma_0$, which lies on the line
$\arg{z}=-\pi/4$ for all $\gamma$. One can show by direct evaluation that
$\lim_{x\to\infty}\erf\left({x \, e^{-i\pi/4}}\right) = 1$ and thus
\begin{align}
	e^{\sigma_0^2 \xi}\left(1 + \erf{\sigma_0 \sqrt{\xi}}\right) \sim 2 e^{\sigma_0^2 \xi}, \label{eq:sigma_0_asymp}
\end{align}
where $\sigma_0^2$ is purely imaginary and the exponent is oscillatory.
For $\gamma > \gamma^{*}$ both $\sigma_1$ and $\sigma_2$ lie on the line $\arg{z}= 3\pi/4$
and by direct evaluation one can show that
$\lim_{x\to\infty}\erf\left({x \, e^{i 3\pi/4}}\right) = -1$.
Similarly to Eq.~\eqref{eq:sigma_2_asymp}, these roots do not contribute to the asymptotic
value and the only important term on big distances is Eq.~\eqref{eq:sigma_0_asymp}.
This results in the following asymptotic pump photon population (red dashed line in Fig.~\ref{fig:1}(b))
\begin{align}
	\lim_{\xi\to\infty}|U(\xi,\gamma)|^2 = \frac{4}{\left(3 + i \gamma/\sigma_0^2\right)^2}.
    \label{eq:pump_at_infinity}
\end{align}

\begin{table}
    \centering
    \begin{tabular}{|c|c|c|c|c|c|}
        \hline
        $\gamma$ & -1.4057 & -2.3981 & -3.0159 & -3.5085 & -3.9330 \\
        \hline
        $\xi_m$ & 1.7895 & 3.6828 & 5.0337 & 6.1353 & 7.0843\\
        \hline
    \end{tabular}
    \caption{First five roots of $I_{p,m}(\gamma)$ and corresponding scaled waveguide lengths $\xi_m$.}
    \label{table:1}
\end{table}
A mathematically equivalent problem of spontaneous emission of an atom with a resonant transition
within photonic band gap was studied in \cite{John:1994-1764:PRA}. Specifically, Laplace image Eq.~\eqref{eq:laplace} is equivalent to Eq.~(2.18) from \cite{John:1994-1764:PRA}, up to a substitution $\Delta\beta_0 \leftrightarrow \delta$, $\alpha \leftrightarrow -i \beta^{3/2}$. 
Since this problem is formulated in time domain, only asymptotic atomic population, equivalent
to Eq.~\eqref{eq:pump_at_infinity}, can be observed experimentally. On the contrary, in our problem
it is possible to adjust the waveguide length to achieve maximum conversion efficiency.
Minimizing numerically $|U(\xi,\gamma)|^2$ with respect to $\xi$ for different values
of $\gamma$ we find optimal scaled waveguide length $\xi_m$
and therefore $z_m = \alpha^{-2/3} \xi_m$ (see Fig.~\ref{fig:1}(a)).
It is clear from Fig.~\ref{fig:1}(b) that for certain values of $\gamma$
the minimum pump wave population $I_{p, m}(\gamma) = |U(\xi_m(\gamma), \gamma)|^2$ becomes zero.
First few such roots are shown in table \ref{table:1}.
Discontinuities in $\xi_m(\gamma)$ observed in Fig.~\ref{fig:1}(a)
and cusps in Fig.~\ref{fig:1}(b) correspond to switching between different
local minima.

\section*{References}
\bibliographystyle{unsrt}
\bibliography{db_NJP_Complete_Photon_Conversion}

\end{document}